# On the nuclear interaction. Potential, binding energy and fusion reaction

I. Casinos

*Facultad de Química, Universidad del País Vasco, Pº Manuel de Lardizabal, 3, 20018 San Sebastián, Spain*

e-mail: ismael.casinos@ehu.es

## Abstract

The nuclear interaction is responsible for keeping neutrons and protons joined in an atomic nucleus. Phenomenological nuclear potentials, fitted to experimental data, allow one to know about the nuclear behaviour with more or less success where quantum mechanics is hard to be used. A nuclear potential is suggested and an expression for the potential energy of two nuclear entities, either nuclei or nucleons, is developed. In order to estimate parameters in this expression, some nucleon additions to nuclei are considered and a model is suggested as a guide of the addition process. Coulomb barrier and energy for the addition of a proton to each one of several nuclei are estimated by taking into account both the nuclear and electrostatic components of energy. Studies on the binding energies of several nuclei and on the fusion reaction of two nuclei are carried out.

# 1 Introduction

A stable atomic nucleus is kept as an entity because the attractive interaction between its component nucleons overcomes the repulsive electric interaction between its protons. The result of both interactions arranges the nucleons in a spherical or spheroid configuration. A nucleus shaped in any of these ways presents a core, whose density is approximately independent on the mass number, and a diffuse surface becoming of zero density [1]. So, the radius of a nucleus is not a precisely defined dimension and its measure depends on the experimental technique used [2].

There is not as yet a unique satisfactory microscopic theory dealing with this attractive nuclear interaction, but potentials for meson-exchange and empirical models intending to account for experimental results [3,4]. Several nuclear potentials and models have been suggested [5,6] to inquire into the nuclear interaction and the nucleonic distribution defining the nuclear structure. The nuclear attractive interaction between nucleons has a short range, similar to the nuclear size. Accordingly, the quantum mechanics formalism should be applied to each nucleus in order to get precise knowledge on it, but difficulties appear to solve the ab initio treatment of a many-body system for non-simple nuclei. Thus, here is used a non-quantum mechanics approach to study $N=Z$-1, $Z$, $Z$+1 nuclei with mass numbers lower than 60 by means of a nuclear potential proposed below.

A similar formalism to that of the gravitational and electrostatic potentials is assumed for this nuclear potential, in spite of some difference in the nature of these universal interactions [7]. On this basis, an expression for the potential energy of two nuclear entities, either nuclei or nucleons, is developed. In order to validate the nuclear potential and estimate parameters in this expression of potential energy, a selection of neutron and proton additions to several nuclei is considered and a model is suggested as a key of the nucleon-addition processes. Coulomb potential barriers and energies for some proton-addition processes are estimated from the respective potential energy profiles by taking into account both the nuclear and electrostatic components of energy. Neutron and proton additions to nuclei have allowed to estimate the binding energies of those $N=Z-1$, $Z$, $Z+1$ nuclei and, on this basis, to study the fusion reaction of two nuclei.

## 2 Potential energy of two nuclear charges

A point located at a distance $r$ from a point nuclear charge $Q$ is assumed to feel an attractive nuclear potential of the form $V = -KQ/r^x$. This starting position is analogous to that of the electrostatic and gravitational potentials, whose known expressions take their own forms by fitting to a value $x = 1$.

In order to find an expression for the nuclear potential felt by a test unit nuclear charge B located at a point B' exterior to an extensive nuclear charge $Q_A$, presenting spherical symmetry (sphere A, in the following) and radius $R_A$, a well-known approach used (see, for example, [8]) to do the same for an electrostatic system is followed here. A homogeneous charge distribution of constant density within the spherical nuclear charge $Q_A$ is assumed in order to integrate the interactions of the unit charge at point B' with every differential volume element constituting the sphere A. In this way, the resultant potential $V$ felt by point B' is expressed by

$$V = \frac{3KQ_A\left\{(r-R_A)^{3-x}\left[r-(x-3)R_A\right]-(r-R_A)^x(r+R_A)^3\left[r+(x-3)R_A\right]\right\}}{2(x-4)(x-3)(x-2)R_A^3 r} \qquad (1)$$

where $r$ is the distance from B' to the sphere centre A'. The nuclear interaction is known to be a short-range one, *i.e.*, it is strongly dependent on $r$, but a satisfactory value for $x$ is not well known and $x = 6$ is assumed, which is below supported by experimental data. In this way, the simple eq. (2) derived from eq. (1) for $x = 6$ is proposed for the nuclear potential outside an extensive nuclear charge $Q_A$

$$V = -KQ_A\Big/\left(r^2 - R_A^2\right)^3 \qquad (2)$$

and the corresponding nuclear field is

$$\Phi = 6KQ_A r\Big/\left(r^2 - R_A^2\right)^4 \hat{\mathbf{r}} \qquad (3)$$

An expression for the potential energy $E$ of the system formed by two extensive nuclear charges $Q_A$ and $Q_B$ presenting spherical symmetry and separated by a distance $r$ between their centres is ascertained by using eq. (2) in an analogous manner to that of above. A homogeneous charge distribution within the spherical nuclear charge $Q_B$ of radius $R_B$ is assumed to integrate the interactions of $Q_A$ with every differential volume element constituting the sphere B leading to

$$E = \frac{3KQ_AQ_B}{8R_B^3 r} \int_0^{R_B} \left\{ \left[ (r+z)^2 - R_A^2 \right]^{-2} - \left[ (r-z)^2 - R_A^2 \right]^{-2} \right\} z \, dz \qquad (4)$$

where $z$ is the distance from the $Q_B$ centre to a differential volume element in this sphere B. This calculation yields the following expression for the potential energy of the spherical nuclear charges $Q_A$ and $Q_B$

$$E = -\frac{3KQ_AQ_B\left(r^2 - R_A^2 - R_B^2\right)}{8R_A^2 R_B^2 \left[ r^2 - (R_A - R_B)^2 \right] \left[ r^2 - (R_A + R_B)^2 \right]} + \frac{3KQ_AQ_B}{32R_A^3 R_B^3} \ln \frac{r^2 - (R_A - R_B)^2}{r^2 - (R_A + R_B)^2} \qquad (5)$$

In order to study a nucleon-nucleus interaction in the following section, a simplifying assumption is made on eq. (4). There can be defined $r^2 - R_A^2 = d^2$ and it is assumed that $r^2$ is higher enough than $R_A^2$ in such a way that $r \approx d$, what occurs when both nuclear charges are separated enough and/or have appreciable different sizes. So, the intended approximate expressions for the potential energy

$$E = -KQ_AQ_B \Big/ \left( r^2 - R_A^2 - R_B^2 \right)^3 \qquad (6)$$

and the corresponding nuclear force

$$\mathbf{F} = 6KQ_AQ_B r \Big/ \left( r^2 - R_A^2 - R_B^2 \right)^4 \hat{\mathbf{r}} \qquad (7)$$

of two nuclear charges $Q_B$ and $Q_B$ are developed.

The equality $r^2 - R_A^2 - R_B^2 = s^2$ can be written for the denominator of eq. (6)

$$E = -KQ_AQ_B \Big/ s^6 \qquad (8)$$

becoming apparent that the interaction of the nuclear charges $Q_A$ and $Q_B$, centred at points A' and B', can be thought as if they were point charges located at points A'' and B'' (effective nuclear interaction centres) along the same line $r$ and separated by a distance $s < r$. A system formed by two non-spherical nuclear charges could not present their mass centres A' and B' and their effective nuclear-interaction centres A'' and B'' on the same line, resulting in an interaction dependent on the mutual orientation. This would cause lower accuracy or inapplicability of eqs. (6) and (7).

## 3 Nucleon-nucleus interaction

According to the precedent section, eq. (6) gives the potential energy of a system formed by two extensive spherical nuclear charges $Q_A$ and $Q_B$ at rest, separated by a distance $r$ between their mass centres. Then, it is assumed that both charges are two nuclei A and B, constituted by $M_A$ and $M_B$ nucleons, respectively, which are structurally arranged according to spherical symmetry. In this way, eq. (6) becomes

$$E = -kM_A M_B \big/ \left( r^2 - R_A^2 - R_B^2 \right)^3 \tag{9}$$

where $Q_A = \alpha\, M_A$, $Q_B = \alpha\, M_B$ and $k = K\, \alpha^2$ are applied, and $\alpha$ is the nuclear charge of one nucleon. Nucleons are treated here as structureless particles.

The process A+n → C represents the addition of a free neutron to a nucleus A, in such a way that both species are initially infinitely separated at rest and such a neutron becomes bonded to the nucleons of nucleus A after a slow mutual approach to constitute the nucleus C. From an opposite point of view, this process represents the separation of the outermost bonded neutron from the nucleus C to yield the nucleus A and a free neutron infinitely separated at rest. The two-body interaction energy between A and n is represented in terms of the potential energy of the individual particles and eq. (9) takes the form

$$S_n(r) = -kM_A \big/ \left( r^2 - R_A^2 - R_n^2 \right)^3 \tag{10}$$

when it is adapted to the characteristics of the process A+n → C.

This addition process is showed in fig. 1 for the illustrative example $^{30}_{15}\text{P} + \text{n} \rightarrow {}^{31}_{15}\text{P}$ by means of the following model. There is an attractive interaction between species A and n from an infinite separation, $r = \infty$, up to a close enough proximity, $r = D_C$, to constitute altogether the nucleus C. However, a repulsive effect appears at $r$ values near $D_C$ that becomes higher than the attractive interaction for $r < D_C$, preventing the nucleus from collapsing toward the centre. As a consequence, an energy minimum is reached for the ground state of nucleus C at an equilibrium distance of radius $D_C$. Along the slow mutual approach of A and n two ranges can be distinguished, which are connected at the point $r = R_A + R_n$, where the spheres A and n, showing more or less diffuse surfaces of radii $R_A$ and $R_n$, get enough proximity to share nucleonic orbital space and therefore affecting the nucleonic distribution of A. So, at $r = R_A + R_n$ is initiated a progressive adjustment of the nucleonic distribution of A and n up to yield the one of C, which occurs in the range $R_A + R_n \geq r \geq D_C$ and concludes at $r = D_C$. Accordingly, the potential energy of the system is given by eq. (10) for $\infty \geq r \geq R_A + R_n$ and it takes the following limit values $S_n(r = \infty) = 0$ and

$$S_n\left(r = R_A + R_n\right) = -kM_A \big/ 8R_A^3 R_n^3 \qquad (11)$$

However, the progressive adjustment of the nucleonic distribution at $R_A + R_n \geq r \geq D_C$, modifying $R_A$ and $R_n$, invalidates the direct applicability of eq. (10) in this range. An expression for $S_n(r)$, applicable in the range $R_A + R_n \geq r \geq D_C$, is derived by assuming that the adjustment of the nucleonic distribution can be modelled by two evolving-in-size spheres keeping close proximity, which undergo the progressive adaptation of their radii to fulfil the limit values $r = R_A + R_n$ and $r = D_C$ of this range. To do this, $R_A$ and $R_n$ are replaced in eq. (11) by the variable radii of the evolving-in-size spheres, $R_1$ and $R_2$, giving rise to

$$S_n(r) = -kM_A \big/ 8R_1^3 R_2^3 \qquad (12)$$

Also, the requirements $R_1 = h_1 r$, $R_2 = h_2 r$ and $h_1 + h_2 = 1$ are suggested to be applied to these evolving spheres in permanent close proximity. By applying the limit condition at point $r = R_A + R_n$ to both radii $R_1$ and $R_2$, that is, $R_A = h_1(R_A + R_n)$ and $R_n = h_2(R_A + R_n)$, then $h_1$ and $h_2$ are expressed as $h_1 = R_A / (R_A + R_n)$ and $h_2 = R_n / (R_A + R_n)$. Finally, by substituting these relations in eq. (12), the intended $S_n(r)$ expression for $R_A + R_n \geq r \geq D_C$ is developed

$$S_n(r) = -\frac{kM_A\left(R_A + R_n\right)^6}{8R_A^3 R_n^3 r^6} \qquad (13)$$

In short, the potential energy $S_n(r)$ along a neutron-addition process A+n → C can be described by eq. (10) for $\infty \geq r \geq R_A + R_n$ and by eq. (13) for $R_A + R_n \geq r \geq D_C$.

In order to ascertain the unknown parameters in eqs. (10) and (13), the following three items are appreciated. First, experimental data of energies for the neutron $S_n$ and proton $S_p$ additions to several nuclei are taken [9]. Second, when eq. (13) is applied to $r = D_C$ becomes

$$S_n = -\frac{kM_A(R_A + R_n)^6}{8R_A^3 R_n^3 D_C^6} \qquad (14)$$

that allows an estimation of the nuclear interaction between the outermost or valence neutron of a nucleus C with the rest of its nucleons, that is, $S_n(r = D_C) = S_n$. Third, the alternating series of the two kinds of nucleon additions A+n → C and A+p → C in the following scheme 1 is written in such a way that a nucleus C formed in a specific addition process is taken as nucleus A that intervenes in the next one

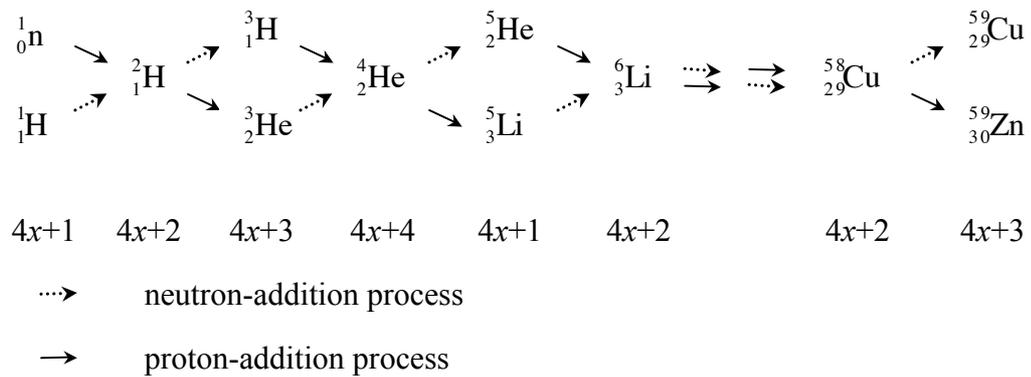

$4x+1 \quad 4x+2 \quad 4x+3 \quad 4x+4 \quad 4x+1 \quad 4x+2 \qquad 4x+2 \quad 4x+3$

   ⋯→     neutron-addition process

   →     proton-addition process

$x = 0, 1, 2, \ldots 14$.

Nuclides in this series are classified in four types according to the calculation of the mass number $M$ either with $4x+1$, $4x+2$, $4x+3$ or $4x+4$ (or $4x$), where $x = 0, 1, 2, \ldots 14$, as indicated in scheme 1. Nuclides with higher $M$ have not been taken because their energy data [9] show lower accuracy or they are unstable.

In order to estimate $D_C$ values in scheme 1, it is assumed that each pair of nuclides, either both of type $4x+1$ or of type $4x+3$, with identical $M$ (i.e., mirror nuclei), has the same value of $D_C$. This can be visualized by considering two mirror nuclei as constituted by two interpenetrated similar networks of neutrons and protons, what suggests that the outermost or valence nucleon, either being neutron or proton, feels a nuclear attraction with the remainder of the nucleons in each network with similar intensity. On the same basis, it is assumed that a nuclide of type $4x+1$ with mass number $M$ has the same $D_C$ value that the subsequent nuclide of type $4x+2$ with $M+1$; and the same assumption is made for a nuclide of type $4x+3$ and the subsequent one of type $4x+4$. These assumptions denote the influence on configuring the nuclear structure by a neutron-proton $np$ pairing of these valence nucleons in both interpenetrated networks. Thus, the difference in two mirror nuclei for the whole interaction of the outermost neutron and that of the outermost proton with the same remainder of nucleons would be only electrostatic (*i.e.,* the difference in the nuclear components of these neutron and proton interactions with the remainder of nucleons is negligible) and it comes expressed by

$$S_p - S_n = K_e e^2 Z_A Z_p / D_C \tag{15}$$

where $Z_A$ is the proton number of nucleus A, and $Z_p = 1$ corresponds to the free proton added to nucleus A to form the nucleus C.

In this way, $D_C$ values for nuclides in scheme 1 (except for deuterium, by an obvious reason) are calculated by using eq. (15) with $S_n$ and $S_p$ data [9] in two pairs of processes: 1) A+n $\rightarrow$ C and A+p $\rightarrow$ C', where C and C' are mirror nuclides either both of type $4x+1$ or of type $4x+3$ with identical $M$, and 2) A+n $\rightarrow$ C and A'+p $\rightarrow$ C, where A and A' are mirror nuclides either both of type $4x+1$ or of type $4x+3$ with identical $M$. Results are shown in fig 2. In fig. 2(a) it can be observed ups and downs in $D_C$ values for alternating $M_C$ values; or, in other words, nuclides C of type $4x+1$ form a subset with the higher $D_C$ values and nuclides C of type $4x+3$ form the other subset with the lower $D_C$ values. Moreover, in fig. 2(b), nuclides C of type $4x+2$ show the same higher $D_C$ values as type $4x+1$, and nuclides C of type $4x+4$ show the same lower $D_C$ values as type $4x+3$. This behaviour is a consequence of the ups and downs shown both by $S_n$ and by $S_p$ data [9] and, because of that, they are shown by the difference $S_p$-$S_n$ in eq. (15). Ups and downs diminish their differences as $M_C$ increases.

Then, it is convenient to express $D_C$ as a function of the mass number. It is known that the approximate constancy of nuclear density gives rise to proportionality between nuclear radius $R$ and mass number according to the relationship $R = R_0 M^{1/3}$, where $R_0$ is a proportionality constant [2,6]. On this basis, each one of the four subsets of $D_C$ values shown in fig. 2 is separately treated with the fit equation

$$D_C = aM_A^{1/3} + b \qquad (16)$$

where $M_A = M_C - 1$ and $a$, $b$ are parameters to be determined from the curve fit. The fit results are presented in table 1.

As indicated above, eq. (14) expresses the energy released in a neutron-addition process $A+n \rightarrow C$. At this point, it is intended to determine the rest of unknown parameters by fitting eq. (14) to $S_n$ data [9] corresponding to each one of the four types of nuclide in scheme 1. To do this, parameters in eq. (14) are first put as a function of $M_A$. The proportionality between $R$ and $M^{1/3}$ considered above to reproduce $D_C$ is now analogously applied to $R_A$ in the form

$$R_A = R_{0A}M_A^{1/3} \qquad (17)$$

where $R_{0A}$ is a proportionality constant to be determined by a curve fit. So, by substituting both eqs. (16) and (17) in eq. (14) it gives rise to

$$S_n = -\frac{k\left(R_{0A}M_A^{1/3} + R_n\right)^6}{8R_{0A}^3R_n^3\left(aM_A^{1/3} + b\right)^6} \qquad (18)$$

where the appropriate $a$ and $b$ values in table 1 are used. However, eq. (18) suffers from the fact that not only one pair of values $R_{0A}$ and $R_n$ fulfil the equation, but any pair that is a multiple of both values. By defining $W = R_{0A}/R_n$ the eq. (18) takes the form

$$S_n = -\frac{k\left(WM_A^{1/3} + 1\right)^6}{8W^3\left(aM_A^{1/3} + b\right)^6} \qquad (19)$$

and the difficulty is surmounted. The fit results are presented in table 2. The fit quality is clearly dependent on the dispersion degree of the experimental $S_n$ points [9]. $W$, or better $W^3 = R_{0A}^3 / R_n^3$, is an estimation of the available space for a bonded nucleon in a nucleus A related to that occupied by a free nucleon.

So, there are found four types of nucleonic distribution in the nuclei of scheme 1, that is, four levels of nuclear density with an associated number of valence nucleons corresponding to the four types of nuclide in scheme 1. A remarkable consequence of this observation is that there is not a single value of $k$ when eq. (9), or other subsequently derived equation, is applied to any pair of interacting nuclei, but neither a different $k$ value is strictly necessary for every pair of interacting nuclei. The single value of $k$ for all the nuclei of the same type in scheme 1 can be understood as if the number of valence nucleons defines the strength of the nuclear interaction, while any number of core nucleons does not change that strength but affects proportionally the overall internuclear interaction. The comparison of the $k$ values for these four types of nuclei displays a dependence of the interaction strength on $np$, $nn$, $pp$ pairings of the valence nucleons as well as a favoured formation of α-particle clusters in nuclei.

On the other hand, the $k$ dependence on the number of valence nucleons in nuclei A and B can be estimated with

$$k = 2TR_{0B}^3 \big/ (R_{0A}^3 + R_{0B}^3) \qquad (20)$$

where $R_{0A}{}^3 / R_{0B}{}^3 \geq 1$, leading to the single and common value of interaction strength $T = 46.695$ MeV fm$^6$ for any pair of interacting nuclei A and B in scheme 1. It is apparent the $k$ dependence on the similarity degree of the available space for one nucleon shown by nuclei A and B according to their respective valence structure. And the potential energy of these nuclei A and B in the range $\infty \geq r \geq R_A + R_B$ can be calculated by means of eq. (9) with appropriate values in table 2 indicated as $R_{0A}$.

Once $W$ is known, it is convenient to ascertain the related $R_{0A}$ and $R_n$ values. In order to find a value for $R_n$, the process p+n $\rightarrow$ $_1^2$H is considered. It can be described as a neutron addition to a proton, which is considered the simplest nuclide A of type $4x+1$, what suggests to use the value $k = 41.129$ MeV fm$^6$ from table 2. The equality of free proton and free neutron radii is assumed, that is, $R_A = R_p = R_n$. And it is also assumed that there is not an adjustment in size for both nucleons from their initial close proximity until their final location constituting a deuteron, that is, $D_C = 2R_n$. These assumptions on the process are applied on eq. (14) leading to $S_n = - k / 8R_n^6$, which yields $R_n = 1.1498$ fm for the experimental value $S_n = - 2.2246$ MeV [9]. This value for the free nucleon radius together with $W$ values in table 2 allow the calculation of $R_{0A}$ corresponding to each type of nuclide by using $R_{0A} = W R_n$. Results are added in table 2 and so all parameters of eqs. (10) and (13) are determined and the potential energy $S_n(r)$ for a process A+n → C of scheme 1 can be calculated.

In fig. 3 is depicted the potential energy along the evolution of the proton and neutron free species up to render a deuteron at $r = D_C$. There can be seen that the interaction weakens up to about 1% at $r = 3.8$ fm, where the nucleonic surfaces are separated 1.5 fm. This short range of the nucleon-nucleon interaction suggests that each nucleon interacts with a limited number of nearest neighbours in more populated nuclei and it is consistent with the higher range of the force for one pion ($\hbar / mc = 1.43$ fm) as carrier of the strong nucleon-nucleon interaction [10].

There have been proposed eq. (10) for $\infty \geq r \geq R_A + R_n$ and eq. (13) for $R_A + R_n \geq r \geq D_C$ to describe the potential energy $S_n(r)$ along a process A+n → C, and the parameters therein have been calculated for each one of the four types of nuclide in scheme 1. It is subsequently intended to propose expressions able to describe the potential energy $S_p(r)$ of a proton-addition process A+p → C. According to the assumptions made above, a proton addition presents the same nuclear interaction component of energy that the corresponding neutron addition, but the former presents as well an electric component contributing to the whole interaction energy between species A and p. Consequently, eq. (21) for $\infty \geq r \geq R_A + R_p$ and eq. (22) for $R_A + R_p \geq r \geq D_C$

$$S_p(r) = -\frac{kM_A}{\left(r^2 - R_A^2 - R_p^2\right)^3} + \frac{K_e e^2 Z_A}{r} \qquad (21)$$

$$S_p(r) = -\frac{kM_A\left(R_A + R_p\right)^6}{8R_A^3 R_p^3 r^6} + \frac{K_e e^2 Z_A}{r} \qquad (22)$$

are suggested to account for the potential energy along the slow mutual approach of the free species A and p from $r = \infty$ until they are bonded, at $r = D_C$, giving rise to nuclide C.

The classification of the processes A+p → C in four subsets, as above, should be taken into account in order to use the appropriate parameter values from tables 1 and 2 for each calculation. Figure 4 shows the potential energy $S_p(r)$ of four processes $^{28}_{14}\text{Si}+\text{p} \rightarrow \,^{29}_{15}\text{P}$, $^{29}_{14}\text{Si}+\text{p} \rightarrow \,^{30}_{15}\text{P}$, $^{30}_{15}\text{P}+\text{p} \rightarrow \,^{31}_{16}\text{S}$, and $^{31}_{15}\text{P}+\text{p} \rightarrow \,^{32}_{16}\text{S}$, where each one is taken as an illustrative example of the corresponding type of nuclide in scheme 1. There are shown the nuclear and electric components of the interaction between A and p as well as the resultant curve $S_p(r)$. In addition, there are pointed out the relative positions of $D_C$, $R_A+R_p$ and Coulomb barrier $CB$ along the evolution of each process. Also, there can be seen that the diffuse surfaces of nuclides A with even number of nucleons present earlier availability to share nucleonic orbital space with the approaching proton than nuclides A with odd number of nucleons to initiate the adjustment of the respective nucleonic distributions.

In a parallel calculation, the Coulomb barriers for the proton-addition processes of scheme 1 are shown in fig. 5. It is remarkable the presence of ups and downs in the $CB$ values for alternating $M_A$ values showing two pairs of subsets: higher values for nuclides A of type $4x+4$ and lower values for nuclides A of type $4x+2$ are shown in fig. 5(a), and higher values for nuclides A of type $4x+1$ and lower values for nuclides A of type $4x+3$ are shown in fig. 5(b). Ups and downs increase their differences as $M_A$ increases.

In other respects, $S_p$ values calculated with eq. (22) for $r = D_C$, by using the appropriate parameters in tables 1 and 2, can be compared with experimental data [9] in fig. 6. There can be seen that nuclides A of type $4x+3$ are those that release the highest energies by adding a proton, what indicates that nuclides C of type $4x+4$ (mass numbers are multiple of 4, the one of the α-particle) present a special stabilization of their nucleonic configuration. And nuclides A of type $4x+4$ present the lowest ability to add a proton to yield nuclides C of type $4x+1$.

## 4 Binding energies

It has been considered the process of a nucleon addition to a nucleus A yielding a nucleus C present in scheme 1. Then, it is intended to compose a nucleus C($Z$, $N$) by the successive and appropriate nucleon-additions following scheme 1 and to estimate the binding energy $E_b(\text{C})$ related to the whole formation process $Z\text{p} + N\text{n}$ $\rightarrow$ C($Z$, $N$). This estimation has been done using eqs. (13) and (22) for $r = D_C$ with the appropriate parameters in tables 1 and 2 for the corresponding nucleon-addition.

Calculated results of binding energies per nucleon $E_b(C)/M_C$ in addition to the corresponding experimental data [9] are presented in fig. 7. Data for mirror nuclides with $N=Z+1$ are showed in fig.7(a) and those for mirror nuclides with $N=Z-1$ in fig.7(b), while nuclides with even $M_C$ are showed in both figures to make easy comparisons. For convenience, plus sign is assigned for $E_b(C)/M_C$ in fig. 7.

The potential energy $E_b(C, r)$ along the previous formation process of a nucleus C is described by considering the synchronous mutual approach of $Z$ protons and $N$ neutrons from all being infinitely separated at rest up to get simultaneously their final positions constituting the nucleonic distribution of C($Z$, $N$). The calculation for $^{59}_{29}$Cu, as an example of C($Z$, $N$), has been done by appropriately applying eqs. (10) and (21) in the range $\infty \geq r \geq R_A + R_n$ and eqs. (13) and (22) in the range $R_A + R_n \geq r \geq D_C$ and it is shown in fig. 8.

## 5 Fusion reactions

The preceding calculation of potential energy $E_b(C, r)$ and binding energy $E_b(C)$ related to the formation process of a nucleus C from its nucleons is now taken as a base to study the fusion reaction A+B $\rightarrow$ C for nuclei present in scheme 1. The exchange in energy $E_f$, or $Q$ value, yielded by this fusion reaction and the potential energy $E_f(r)$ along the process are explicit in the following expressions

$$E_f = E_b(C) - E_b(A) - E_b(B) \tag{23}$$

$$E_f(r) = E_b(C, r) - E_b(A, r) - E_b(B, r) \tag{24}$$

and all these terms can be calculated as indicated above. When the point $r = D_C$ is considered in eq. (24), each term leads to the corresponding term in eq. (23), where the symbol $D_C$ is not written for convenience.

Figure 9 exemplary shows the potential energy of the nuclei A (*e.g.*, $^{32}_{16}$S) and B (*e.g.*, $^{27}_{13}$Al) along the mutual approach from an infinite separation up to get fused constituting the nucleus C (*e.g.*, $^{59}_{29}$Cu). There is appreciated a potential energy maximum to be surpassed by the nuclei A and B to yield C at an interaction or fusion barrier radius $R_i$.

This model considers the fusion process as the synchronous occurrence of two gradual events: decomposition of nucleus B and composition of nucleus C by appropriately distributing the nucleons from B onto nucleus A. The first event is given by the term $E_b(B, r)$ in eq. (24) and the second event is given by the difference $E_b(C, r) - E_b(A, r)$ in the same equation.

Three critical points along the fusion process in fig. 9 have the following meaning: 1) at $r = R_A + R_B = R_{AB}$ the surface diffuseness radii of nuclei A and B reach an external contact and their nucleonic distributions start to be affected, 2) at $r = D_A + D_B = D_{AB}$ the nuclei A and B proceed to exchange mass by forming a partially fused composite or dinuclear system DNS, and 3) at $r = D_C$ the mass exchange in DNS is finished and the compound nucleus CN of nucleus C is formed. However, while this mass exchange takes place in a more or less asynchronous manner by forming a neck in DNS this model considers the mass evolution in a synchronous way and because of that an energy term $E_a(R_i)$ is suggested below to take into account this dynamic effect.

In order to examine the fusion cross-section $\sigma_f$ as a function of the incident energy in the centre-of-mass system $E_{c.m.}$, the conservation of total energy is applied to the potential energy maximum where $r = R_i$

$$E_{c.m.} = E_f(R_i) + E_{cent}(R_i) + E_a(R_i) + E_{rv}(R_i) \qquad (25)$$

$E_f(R_i)$ is calculated with eq. (24) for $r = R_i$. The centrifugal energy is estimated for $r = R_i$ with

$$E_{cent}(r) = t(r)\, L^2 / 2\mu\, r^2 \qquad (26)$$

where $L^2 = 2\mu\, E_{c.m.} b^2$; $\mu,\ L,\ b$ are reduced mass, angular momentum and impact parameter, respectively; and the DNS evolution factor

$$t(r) = \frac{5 M_A M_B (D_{AB} - r) + 2 M_C^2 (r - D_C)}{2 M_C^2 (D_{AB} - D_C)} \qquad (27)$$

allows to give an account of the mass-exchange progress reached by DNS at $r = R_i$. Complementarily, where the nuclei A and B have not started the DNS formation is $t(r \geq D_{AB}) = 1$ and where CN is accomplished and one refers to rotational energy is $t(r = D_C) = 5 M_A M_B \big/ 2 M_C^2$. An estimation of the reduction in the synchronous potential energy at $r = R_i$, as a consequence of the asynchronous character of the evolution A+B $\rightarrow$ DNS $\rightarrow$ CN, is taken into account by

$$E_a(R_i) = \frac{R_i - R_{AB}}{R_{AB}} [E_{fc}(R_i) - E_{fc}(R_{AB})] \tag{28}$$

where $E_{fc}(R_i)$ and $E_{fc}(R_{AB})$ are effective potential energies calculated with $E_{fc}(r) = E_f(r) + E_{cent}(r)$ for $r = R_i$ and $r = R_{AB}$. A radial extra-push energy $E_{rv}(R_i)$ appearing at high enough $E_{c.m.}$ is needed to achieve that DNS can become CN and/or to avoid any other competitive process, as deep inelastic collision or quasi-fission. This is realized by complementing the centrifugal energy in DNS up to the rotational energy in CN at constant $L$ for systems with $1 < M_A/M_B < 4$ and/or by dissipating an appropriate fraction of the kinetic energy into internal excitation energy by means of a high-enough frictional collision between the nuclei A and B which is undertaken with a low-enough impact parameter. But an expression for $E_{rv}(R_i)$ has not been attained at present.

Thus, the fusion cross-section is described by taking $b^2$ from $E_{cent}(R_i)$ of eq. (25) into the sharp cutoff relationship $\sigma_F = 10\pi b^2$ to render the form

$$\sigma_f = \frac{10\pi R_i^2}{t(R_i)} \left[ 1 - \frac{E_f(R_i) + E_a(R_i) + E_{rv}(R_i)}{E_{c.m.}} \right] \tag{29}$$

where $\sigma_F$, $R_i$ and $E$ are expressed in mb, fm and MeV, respectively.

The model is tested with two fusion reactions $^{32}_{16}\text{S} + ^{27}_{13}\text{Al}$ and $^{19}_{9}\text{F} + ^{40}_{20}\text{Ca}$ which lead to the same nucleus $^{59}_{29}\text{Cu}$. Cross-section data for these systems are taken from refs. [11-16] and they are shown in fig 10. There are appreciated the regions I, II and III related to the three ranges of low, medium and high $E_{c.m.}$, respectively. Four characteristic points $P_{1-4}$ are also indicated: $P_1$ denotes the lowest $E_{c.m.}$ able to yield fusion in a head-on collision, $P_2$ and $P_3$ denote the limits between regions I-II and between regions II-III, respectively, and $P_4$ denotes the highest entrance channel leading to fusion.

It is suggested that the radial extra-push energy $E_{rv}(R_i)$ is zero in region I, that it starts slowly and determines region II, and that it increases strongly and determines region III. As a consequence, the impact parameter and accordingly the fusion cross-section increase in region I, and they decline slightly in region II and strongly in region III. However, not only $b$ and $\sigma_f$ decrease in region III by increasing $E_{c.m.}$, the point $P_4$ is reached where the angular momentum begins to decrease and its maximum $L_{max}$ occurs there. The $\sigma_f$ behaviour in fig. 10 is related to $L^2$ by the slope of each straight line passing through the coordinates origin according to $\sigma_f = 5\pi L^2/\mu\ E_{c.m.}$ (from $L^2 = 2\mu\ E_{c.m.}b^2$ and $\sigma_f = 10\pi b^2$) which denotes two $\sigma_f$ points, except the $L^2_{max}$ line that denotes only the point $P_4$.

An intended general way to calculate $\sigma_f$ as a function of $E_{c.m.}$ by estimating all the energy terms of eq. (29) in the range $0 \leq L \leq L_{max}$ is obstructed by the lack of an expression for $E_{rv}(R_i)$. Consequently, the $\sigma_f$ calculation has been made in region I, but in regions II and III, where $E_{rv}(R_i) \neq 0$, it is only made for the angular momentum values corresponding to the characteristic points $P_{2-4}$. In this way, fig. 11 shows the effective potential energy $E_{fe}(r)$ as a function of the internuclear distance $r$ for the four selected entrance channels in the range $0 \leq L \leq L_{max}$ corresponding to the points $P_{1-4}$. A pocket appears because any $E_{fe}(D_C)$ value becomes lower than the corresponding $E_{fe}(R_i)$ one. The location of the points $r = R_{AB}$, $D_{AB}$, $D_C$ is indicated and the $R_i$ position decreases from 6.98 to 6.30 fm by increasing the angular momentum from $L_1$ to $L_4 = L_{max}$. The angular momentum values $L_2$, $L_3$ and $L_4$, corresponding to the characteristic points $P_2$, $P_3$ and $P_4$, are estimated by applying suitable requirements to the appropriate energy terms as indicated below. And the energy terms depending on these $L_{2-4}$ values are then calculated as above in order to know $E_{c.m.}$ with eq. (25), and subsequently $b^2$ and $\sigma_f$ are calculated with $L^2 = 2\mu\ E_{c.m.}b^2$ and $\sigma_f = 10\pi b^2$. So, $L_2$ is estimated with the requirement $E_{fe}(D_C) = E_{fe}(R_i) - E_{fe}(R_{AB})$ and the related $E_{c.m.}$ is evaluated by taking $E_{rv}(R_i) = 0$; $L_3$ is estimated with the requirement $E_{fe}(D_C) = 2E_{fe}(R_{AB})$ and the related $E_{c.m.}$ is evaluated by becoming $E_{rv}(R_i) = E_{cent}(R_{AB})$; and $L_4$ is estimated with the requirement $E_a(R_i) = -E_f(R_i)$ (when the highest effect of the asynchronous character of the evolution A+B → DNS → CN is reached) and the related $E_{c.m.}$ is evaluated by becoming $E_{rv}(R_i) = E_{fe}(R_i)$.

And several $L$ values in the range $0 \leq L \leq L_2$ of region I are considered to evaluate the related $E_{c.m.}$ values with eq. (25) by taking $E_{rv}(R_i) = 0$.

Finally, $\sigma_f$ values estimated for the characteristic points P$_{2-4}$ and for the points in region I are added together the experimental data in fig. 10.

In conclusion, a fairly good agreement between calculated and experimental data denotes a suitability of the equations and parameters used in these calculations as well as of the diverse assumptions made and the model suggested.

I am grateful to I. Tellería for valuable help.

**Table 1.** Parameters $a$ and $b$ fitted with eq. (16).

| Type of nuclide A | Type of nuclide C | $a$ (fm) | $b$ (fm) | Correlation coefficient |
|---|---|---|---|---|
| 4$x$+1 | 4$x$+2 | 0.71841 | 1.3948 | 0.97707 |
| 4$x$+2 | 4$x$+3 | 0.87530 | 0.82888 | 0.99225 |
| 4$x$+3 | 4$x$+4 | 0.92065 | 0.64997 | 0.99092 |
| 4$x$+4 | 4$x$+1 | 0.68994 | 1.5116 | 0.97544 |

**Table 2.** Parameters $k$ and $W$ fitted with eq. (19). $R_{0A}$ is calculated with $R_{0A} = W R_n$.

| Type of nuclide A | $k$ (MeV fm$^6$) | $W$ | Correlation coefficient | $R_{0A}$ (fm) |
|---|---|---|---|---|
| 4$x$+1 | 41.129 | 1.0313 | 0.96109 | 1.1858 |
| 4$x$+2 | 11.778 | 1.9513 | 0.92935 | 2.2436 |
| 4$x$+3 | 43.113 | 1.1108 | 0.85601 | 1.2772 |
| 4$x$+4 | 4.8798 | 2.5457 | 0.95406 | 2.9271 |

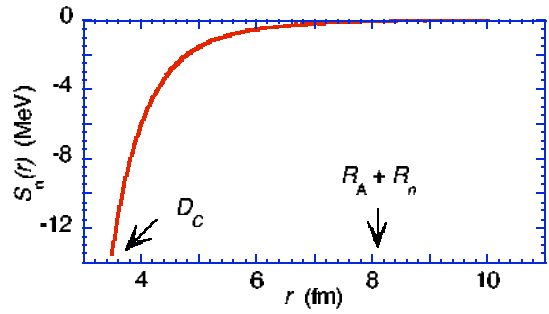

**Fig. 1.** Illustrative plot of the potential energy of a neutron addition to a nucleus (*e.g.*, $^{30}_{15}\text{P} + \text{n} \rightarrow \, ^{31}_{15}\text{P}$).

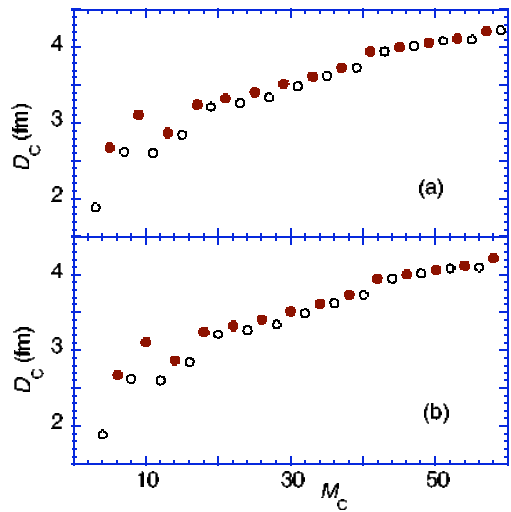

**Fig. 2.** $D_C$ values calculated for nuclides of scheme 1. (a), ups and downs of $D_C$ values for alternating $M_C = 4x+1$ (filled circles) and $M_C = 4x+3$ (unfilled circles) nuclides. (b), ups and downs of $D_C$ values for alternating $M_C = 4x+2$ (filled circles) and $M_C = 4x+4$ (unfilled circles) nuclides.

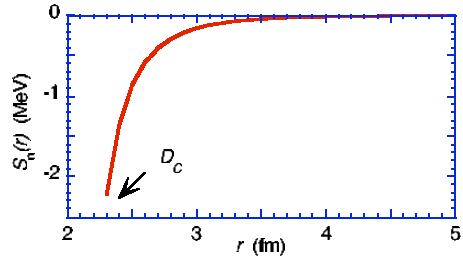

**Fig. 3.** Potential energy calculated along the proton and neutron evolution up to yield a deuteron.

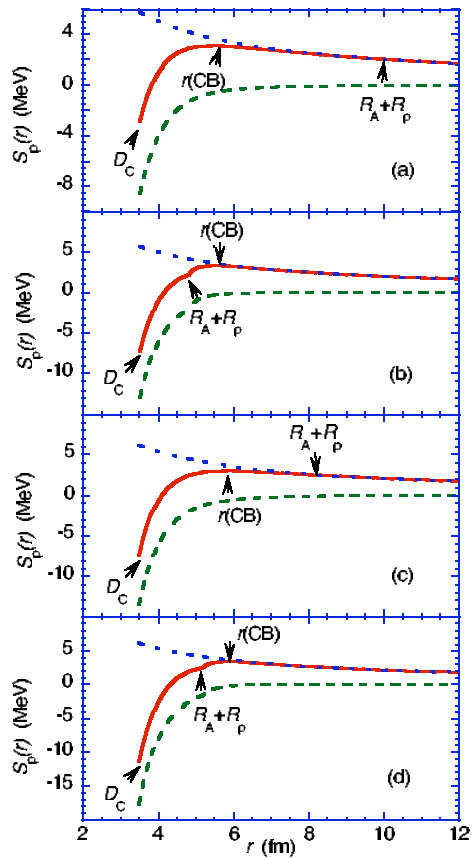

**Fig. 4.** Potential energy calculated along a proton-addition process. Illustrative processes of four types of nuclide $4x+4$, $4x+1$, $4x+2$ and $4x+3$ are shown: (a), $^{28}_{14}\text{Si}+\text{p} \rightarrow {}^{29}_{15}\text{P}$; (b), $^{29}_{14}\text{Si}+\text{p} \rightarrow {}^{30}_{15}\text{P}$; (c), $^{30}_{15}\text{P}+\text{p} \rightarrow {}^{31}_{16}\text{S}$ and (d), $^{31}_{15}\text{P}+\text{p} \rightarrow {}^{32}_{16}\text{S}$. The nuclear (dashed line) and electric (dotted line) components of the resultant $S_\text{p}(r)$ curve (solid line) are all shown.

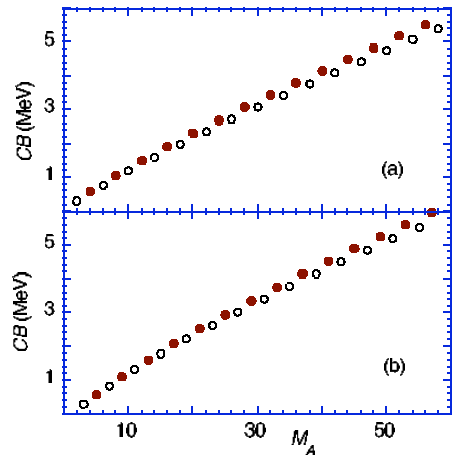

**Fig. 5.** Calculated Coulomb barriers for proton-addition processes indicated in scheme 1. (a), ups and downs of *CB* values for alternating $M_A = 4x+4$ (filled circles) and $M_A = 4x+2$ (unfilled circles) nuclides. (b), ups and downs of *CB* values for alternating $M_A = 4x+1$ (filled circles) and $M_A = 4x+3$ (unfilled circles) nuclides.

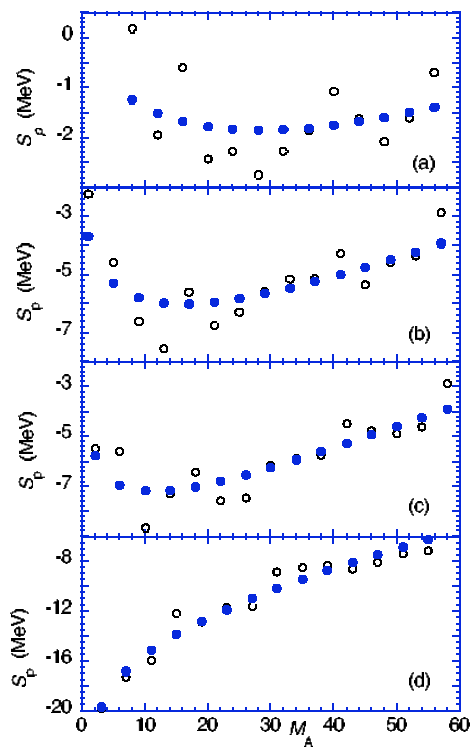

**Fig. 6.** Experimental (unfilled circles) and calculated (filled circles) proton-addition energies for the following types of nuclide: (a), $M_A = 4x+4$; (b), $M_A = 4x+1$; (c), $M_A = 4x+2$ and (d), $M_A = 4x+3$.

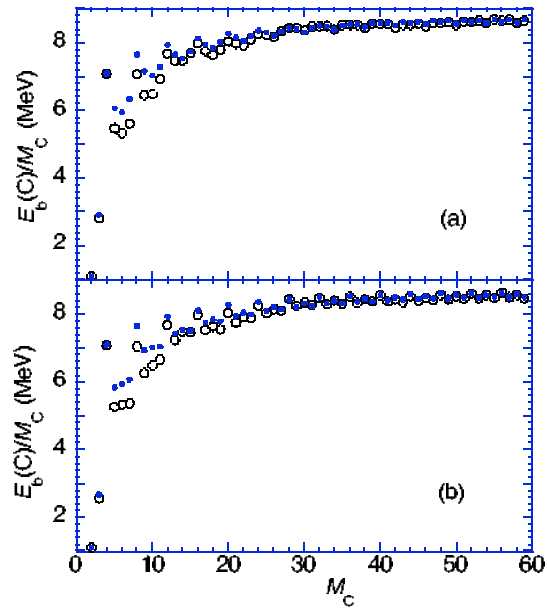

**Fig. 7.** Experimental (unfilled circles) and calculated (filled circles) binding energies per nucleon for nuclides in scheme 1. (a), odd $M_C$ nuclides with $N = Z+1$ and even $M_C$ nuclides. (b), odd $M_C$ nuclides with $N = Z-1$ and even $M_C$ nuclides.

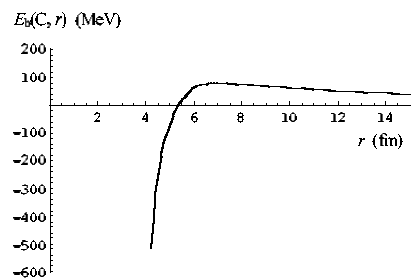

**Fig. 8.** Potential energy along the synchronous evolution of $Z$ protons and $N$ neutrons up to yield the nucleus $C(Z, N)$ (*e.g.*, $^{59}_{29}$Cu).

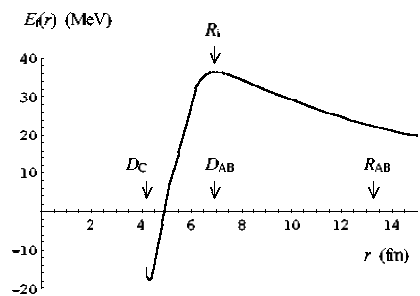

**Fig. 9.** Potential energy along the synchronous evolution of nuclei A and B up to yield the nucleus C (*e.g.*, $^{32}_{16}$S + $^{27}_{13}$Al → $^{59}_{29}$Cu ).

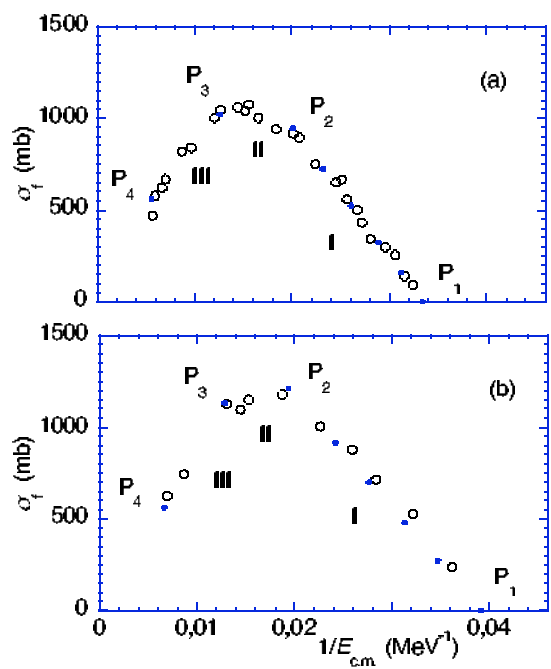

**Fig. 10.** Experimental (unfilled circles) and calculated (filled circles) cross-sections for the fusion reactions: (a), $^{32}_{16}$S + $^{27}_{13}$Al → $^{59}_{29}$Cu and (b), $^{19}_{9}$F + $^{40}_{20}$Ca → $^{59}_{29}$Cu .

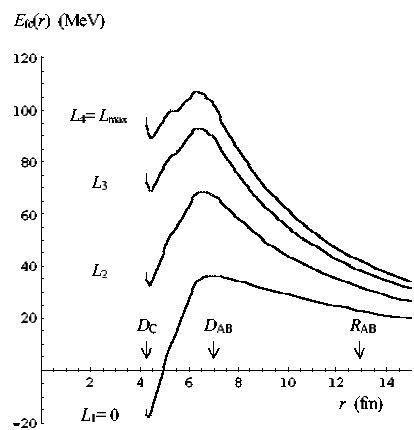

**Fig. 11.** Effective potential energies along the fusion process $^{32}_{16}S + \, ^{27}_{13}Al \rightarrow \, ^{59}_{29}Cu$ for the values $L_{1\text{-}4}$ of angular momentum related to the characteristic points $P_{1\text{-}4}$.